\documentstyle[twocolumn,eqsecnum,aps,epsfig]{revtex}
\newcommand{\be}{\begin{equation}}
\newcommand{\ee}{\end{equation}}
\newcommand{\bea}{\begin{eqnarray}}
\newcommand{\eea}{\end{eqnarray}}

\def\Journal#1#2#3#4{{#1} {\bf #2}, #3 (#4)}

\def\NP{{Nucl. Phys.}}
\def\PL{{Phys. Lett.} B}
\def\PR{{Phys. Rept.}}
\def\PRL{Phys. Rev. Lett.}
\def\PRD{{Phys. Rev.} D}
\def\PRC{{Phys. Rev.} C}
\def\ZPA{{Z. Phys.} A}
\def\ZPC{{Z. Phys.} C}
\def\JPG{{J. Phys.} G}

\def\MPLA{{Mod. Phys. Lett. A}}

\def\EPJ{{Eur. Phys. J.}}

\begin{document}
\title{Direct emission of multiple strange baryons in ultrarelativistic
heavy-ion collisions from the phase boundary}
\author{A.\ Dumitru$^1$, S.A.\ Bass$^2$, M.\ Bleicher$^3$,
H.\ St\"ocker$^3$, W.\ Greiner$^3$}
\address{$^1$Physics Department, Yale University,
P.O.\ Box 208124, New Haven, CT 06520, USA\\
$^2$Department of Physics, Duke University,
Durham, NC 27708, USA\\
$^3$Institut f\"ur Theoretische Physik, J.W.\ Goethe Universit\"at,
Robert-Mayer Str.\ 10, 60054 Frankfurt, Germany}
\maketitle   
\begin{abstract}
We discuss a model for the space-time evolution of ultrarelativistic heavy-ion
collisions which employs relativistic hydrodynamics within one region
of the forward light-cone, and microscopic transport theory (i.e.\
UrQMD) in the complement. 
Our initial condition consists of a quark-gluon plasma
which expands hydrodynamically and hadronizes.
After hadronization the solution
eventually changes from expansion in local equilibrium to
free streaming, as determined selfconsistently by the interaction rates
between the hadrons and the local expansion rate.
We show that in such a scenario the inverse slopes of the $m_T$-spectra
of multiple strange baryons ($\Xi$, $\Omega$)
are practically unaffected by the purely
hadronic stage of the reaction, while the flow of $p$'s and
$\Lambda$'s increases. Moreover, we find that the rather ``soft'' transverse
expansion at RHIC energies (due to a first-order
phase transition) is not washed out by strong rescattering in the hadronic
stage. The earlier kinetic freeze-out as compared to SPS-energies results in
similar inverse slopes (of the $m_T$-spectra of the hadrons in the final state)
at RHIC and SPS energies.
\end{abstract}
\narrowtext

Ultrarelativistic heavy ion collisions offer the unique opportunity to
study highly excited QCD-matter in the laboratory, and possibly the QCD phase
transition to the so-called quark-gluon plasma (QGP) at high energy
density~\cite{QGP}. The dynamics of such reactions is commonly
described either within hydrodynamics or within microscopic
transport models~\cite{StoePR}. The fluid-dynamical approach is most
accurate in very dense systems (where mean free paths are small) close
to local equilibrium. In this model the reaction dynamics is closely
linked to the equation of state (EoS),
which enters directly into the equations of motion. It is therefore well
suited to study systems that undergo phase transitions, e.g.\ that of
hadronic matter into a QGP. In particular, one can employ an EoS
with a first order phase transition, where the QGP coexists with hadronic
matter within some region of energy- and baryon density.

Hadronic cascades based on binary hadron-hadron collisions are probably
less well suited to study very dense systems, and of hadrons
coexisting with a QGP. However, they are useful for the dilute
space-time regions since they account for finite relaxation times in
the hadron gas~\cite{relax}
and the breakup of local kinetic equilibrium. Moreover, since each hadron
is propagated individually, and its interactions with other hadrons
are described on the basis of elementary processes, microscopic transport
models offer the opportunity to {\em calculate} the freeze-out conditions
instead of just putting them in by hand as is usually done in fluid-dynamical
approaches.
In particular, one needs not assume that
all hadron species decouple on the same
hypersurface~\cite{microFO,HKM}.

The sequencial freeze-out is particularly important to understand~\cite{thOm}
the experimental fact that in Pb+Pb collisions at CERN-SPS energy
($\sqrt{s}=17A$~GeV) the transverse mass spectra of multiple
strange baryons~\cite{expOm,NA49chi} are ``softer'' than expected by
a linear interpolation of the inverse slopes of protons and
deuterons~\cite{Tmsyst}: If all hadrons flow with
the same collective velocity and decouple on the same hypersurface,
the inverse slopes or average transverse momenta have to increase with the
mass of the hadron (at least if one neglects resonance decays).

We shall discuss here, whether it is possible to understand these observations
by assuming the formation of a QGP which expands as an ideal relativistic
fluid, going through a mixed phase where it coexists with hadronic matter.
After hadronization is completed (locally in space-time), we evolve
the system within microscopic transport theory (we employ the {\small UrQMD}
model~\cite{UrQMD}), and follow the evolution of the hadrons until
they freeze-out. We point out that the hadrons are ``born'' in an
expanding three-volume (i.e.\ the hadronization hypersurface), the expansion
rate of that volume being many orders of magnitude larger than in
the universe at hadronization~\cite{ExpSc}. It will thus depend on the
corresponding elementary cross-sections and the composition of the
system which hadrons are able to maintain local kinetic equilibrium,
despite the very large expansion rate.
We will show in particular how the momentum distributions
of various hadron species are affected by the hadronic stage
as compared to those on the hadronization hypersurface.

Let us first briefly describe the specific form of hydrodynamics that we shall
employ. For a more detailed discussion see~\cite{Bj,DumRi}.
For simplicity, we assume boost-invariant longitudinal flow,
$v_z=z/t$. This should be a reasonable first approximation
around midrapidity. Cylindrically
symmetric transverse expansion is superimposed.
For $T>T_C=160$~MeV we employ the well-known MIT bagmodel equation of state,
assuming for simplicity an ideal gas of quarks,
antiquarks (with masses $m_u=m_d=0$, $m_s=150$~MeV), and gluons.
For $T<T_C$ we employ an ideal hadron gas that includes the complete
hadronic spectrum up to a mass of 2~GeV.
At $T=T_C$ we require that both pressures are equal, which
fixes the bag constant to $B=380$~MeV/fm$^3$.
By construction the EoS exhibits a first-order phase transition.

For collisions at SPS energy we assume that hydrodynamic flow sets in
on the proper time hyperbola $\tau_i=1$~fm/c. This is a value conventionally
assumed in the literature, cf.\ e.g.~\cite{Bj,DumRi}.
We further employ a (net) baryon rapidity density (at midrapidity) of
$dN_B/dy=80$, as obtained by the NA49-collaboration for central Pb+Pb
reactions~\cite{NA49netB}. The average specific entropy in these collisions is
$\overline s/\overline\rho_B=45\pm5$ (the bar means averaging over the
transverse plane). With this entropy per net baryon most measured hadron
multiplicity ratios can be described within $\pm20\%$~\cite{BMS}.
The corresponding initial energy and net baryon densities
($\overline\epsilon_i=6.1$~GeV/fm$^3$, $\overline\rho_i=4.5\rho_0$)
are assumed to be distributed in the transverse
plane according to a so-called ``wounded nucleon'' distribution
with transverse radius $R_T=6$~fm~\cite{DumRi}.

The measured $p_T$-spectra (at midrapidity) of $\pi$'s, $p$'s, $\Lambda$'s,
$\phi$'s, and $d$'s can be reproduced within this model
if one assumes an ideal hydrodynamic
expansion between the $\tau=\tau_i$ hyperbola and the $T=130$~MeV
isotherm~\cite{DumRi}.
However, the resulting spectra of $\Xi$'s and $\Omega$'s are
{\em harder} than those of $p$'s and $\Lambda$'s,
because the same freeze-out hypersurface is
employed for all hadrons.

Due to the higher parton
density at midrapidity (as compared to collisions at SPS energy),
thermalization may be reached earlier at RHIC. According to various
studies~\cite{T0}, it may occur within
$0.2-1$~fm. We employ $\tau_i=0.6$~fm.
The net baryon rapidity density and specific entropy
at midrapidity in central Au+Au at $\sqrt{s}=200A$~GeV is predicted by
various models of the initial evolution
to be in the range $dN_B/dy\approx20-35$, $s/\rho_B\approx
150-250$~\cite{microRHIC}. We will employ
$dN_B/dy=25$ and $\overline s/\overline\rho_B=205$
($\rightarrow\overline\epsilon_i=20$~GeV/fm$^3$,
$\overline\rho_i=2.3\rho_0$).
(The initial conditions could of course be fine-tuned once the first
experimental data are available.)
This yields a transverse energy on the $\tau=\tau_i$ hyperbola
of $dE_T/dy=1.3$~TeV, which decreases to 720~GeV on the
hadronization hypersurface. We find that $dE_T/dy$ is almost conserved during
the following hadronic evolution, i.e.\ dissipation compensates for the
expansion work.

Having specified the initial conditions and the EoS, the hydrodynamical
solution between the $\tau=\tau_i$ hyperbola and the hadronization
hypersurface is uniquely determined. For a more detailed discussion of
how this solution is obtained, cf.~\cite{DumRi}.

The number of hadrons of species $i$ hadronizing at space-time rapidity
$\eta$, proper time $\tau$, and
position $r_T\left(\sin\left(\chi-\phi\right),\cos\left(\chi-\phi\right)
\right)$,
with four-momentum $p^\mu=(m_T\cosh y,p_T\sin\chi,
p_T\cos\chi,m_T\sinh y)$, is given by~\cite{CF}
\bea 
\frac{dN_i}{d^2m_Tdyd\eta d\zeta d\phi} &=&
r_T\,\tau\left( p_T\cos\phi
\frac{d\tau}{d\zeta} \right.\nonumber\\
&-& \left. m_T\cosh(y-\eta)\frac{dr_T}{d\zeta}\right)\,
f\left(p\cdot u\right),\label{CFspectra}
\eea
where $u^\mu=\gamma_T (\cosh\eta,v_T\sin(\chi-\phi),v_T\cos(\chi-\phi),
\sinh\eta)$ denotes the fluid four-velocity.
$\zeta\in [0,1]$ parametrizes the hadronization
hypersurface\footnote{The switch is actually done slightly behind the
hadronization hypersurface, cf.\ the discussion in~\cite{DumRi}, because
the rate of expansion of its volume can diverge~\cite{ExpSc}.}
(counter-clockwise), such that $r_T(\zeta)$ and $\tau(\zeta)$
specify the space-time points on the hypersurface.
$f$ is either a Bose-Einstein or Fermi-Dirac distribution function. We 
assume that $v_T$, $T$,
and the chemical potentials, are independent of $\eta$,
such that the hadronization hypersurface in the
$t-z$ plane (fixed $r_T$) is simply a proper-time hyperbola.

The ensemble of hadrons generated according to eq.~(\ref{CFspectra}) is
now taken as input for the microscopic transport model {\small UrQMD}.
Thus, after formation
each hadron is propagated individually along a classical trajectory
and can scatter stochastically
until it eventually does not interact any more and freezes out.
Note that by construction the hadron fluid starts from a state of
local equilibrium, but in a three-space with very rapidly increasing
volume-measure~\cite{ExpSc}. The energy-momentum tensors and conserved
currents (of UrQMD and hydrodynamics) match on the hadronization hypersurface,
because the local collective flow velocities, as well as the energy
and net baryon densities, and the pressures, are equal (we include the
same states in the hadronic part of the fluid-dynamical EoS as in UrQMD).

The {\small UrQMD} model
is based on the covariant propagation of all hadrons,
excitation of resonances and strings and their
subsequent decay resp.\ fragmentation.
Free cross-sections for hadron-hadron
scattering are employed. Comparisons of 
{\small UrQMD} calculations to various experimental data from
SIS to SPS energies, as well as a detailed description of the model,
are documented elsewhere~\cite{UrQMD}.

\begin{figure}[htp]
\vspace*{-.5cm}
\centerline{\hbox{\epsfig{figure=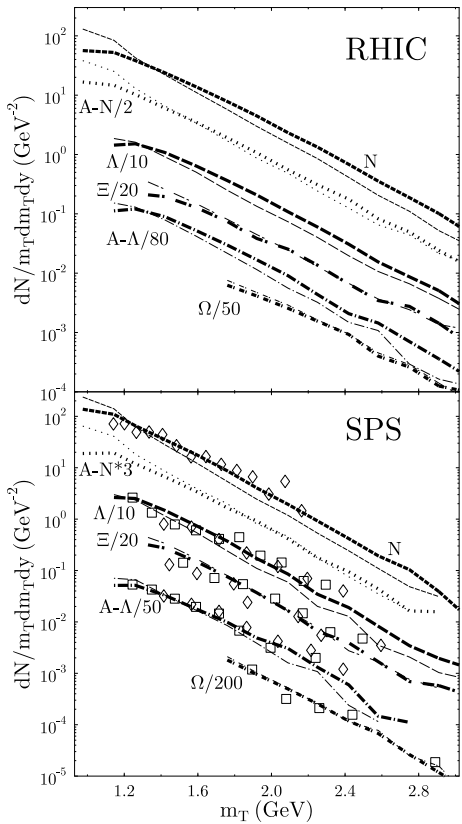,width=6cm}}}
\caption{Transverse mass spectra of $N$, $\overline{N}$, $\Lambda+\Sigma^0$,
$\Xi^0+\Xi^-$, and $\Omega^-$ at $y_{c.m.}=0$ in central Pb+Pb collisions at
$\protect\sqrt{s}=17A$~GeV (lower panel) and Au+Au at
$\protect\sqrt{s}=200A$~GeV (upper panel).
Thin lines: on the hadronization hypersurface; thick lines: at freeze-out;
diamonds: NA49 data; squares: WA97 data.}
\label{sps_mt}
\end{figure}

Fig.~\ref{sps_mt} compares the $m_T$-spectra on the hadronization
hypersurface (thin lines), obtained from Eq.~(\ref{CFspectra}) (plus
strong resonance decays), with those at freeze-out (thick lines). The
$(+)$-$(-)$~\cite{NA49plmi}, $\Lambda$, $\overline\Lambda$~\cite{NA49L},
and $\Xi$~\cite{NA49chi}
spectra of NA49, and the $\Lambda$, $\overline\Lambda$, $\Xi$, and
$\Omega$ spectra of WA97~\cite{expOm} are also depicted (the latter have
been normalized to our calculation). Clearly, because of rescattering,
the transverse flow of $N$'s and $\Lambda$'s increases during
the hadronic stage. Also, slow antinucleons (with
$m_T<1.5$~GeV) are annihilated to a large extent because the
annihilation cross-section increases rapidly at small momenta.
On the other hand, the spectra of $\Omega$'s and of $\Xi$'s with
$m_T\stackrel{>}{~}1.6$~GeV are practically
unaffected by the hadronic stage and closely resemble those on the phase
boundary. 
This is due to the fact that the scattering rates of $\Xi$ and
$\Omega$ in a pion-rich hadron gas are smaller than those
of $N$'s and $\Lambda$'s~\cite{thOm}.
Note that a $\pi$ and $\Omega$ can not be coupled to any of the
known~\cite{PDg} $B=1$, $S=-3$ states
with masses $<2.5$~GeV. The excitation
of a state above this threshold requires a large ($\Delta E\sim1$~GeV)
energy transfer from the $\pi$ to the $\Omega$ (in the local rest-frame),
which is highly suppressed
since on the hadronization hypersurface the $\pi$-energies are distributed
thermally with $\langle E\rangle_\pi\approx 3T_C$.
Finally, elastic collisions give much smaller
energy and momentum transfers to the $\Omega$
than $m_\Omega$, and therefore do not change its $m_T$-distribution.

\begin{figure}[htp]
\vspace*{-.5cm}
\centerline{\hspace{.8cm}\hbox{\epsfig{figure=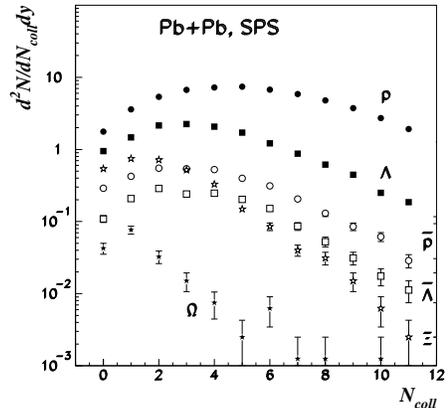,width=6cm}}}
\caption{Distribution of the number of interactions that the final-state
particles suffer after being hadronized.}
\label{ncoll}
\end{figure}

As shown in Fig.~\ref{ncoll}, on average
the baryons which finally emerge as $\Xi$'s and $\Omega$'s suffer
even less interactions than the final-state $\overline{p}$'s and
$\overline{\Lambda}$'s. Thus, within the model presented here,
these particles are emitted {\em directly from the hadronization hypersurface}
without further rescattering in the hadronic stage.
The hadron gas emerging from the hadronization of the QGP (in these
high-energy reactions) is almost ``transparent'' for the multiple
strange baryons, especially because of the large expansion rate. 
In our calculation, the flow seen in the $\Omega$-spectra at SPS and RHIC
is fully accounted for by the expansion preceeding hadronization
(for entropy and baryon density as discussed above).
The space-time domains of freeze-out
for several hadron species are discussed in~\cite{BassDu}.

\begin{figure}[htp]
\vspace*{-.5cm}
\centerline{\hspace{.8cm}\hbox{\epsfig{figure=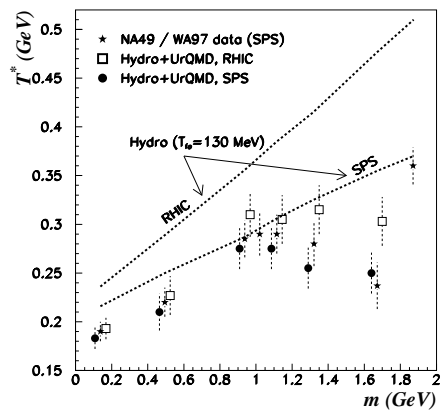,width=6cm}}}
\caption{Inverse slopes of the $m_T$-spectra of $\pi$, $K$, $p$,
$\Lambda+\Sigma^0$, $\Xi^0+\Xi^-$, and $\Omega^-$ at $y_{c.m.}=0$,
$m_T-m_i<1$~GeV.}
\label{slopes}
\end{figure}

Fig.~\ref{slopes} displays the inverse slopes $T^*$ obtained by a fit of
$dN_i/d^2m_Tdy$ to $\exp(-m_T/T^*)$ in the range $m_T-m_i<1$~GeV (for
clarity, the calculated points have been slightly displaced to the left or
right of the true hadron mass).
The trend of the SPS data, namely the ``softer'' spectra of $\Xi$'s and
$\Omega$'s as compared to a linear $T^*(m)$ relation, is reproduced
reasonably well. This is in contrast to ``pure'' hydrodynamics with kinetic
freeze-out on a common hypersurface (e.g.\ the $T=130$~MeV isotherm), where
the stiffness of the spectra increases with mass, cf.\
Fig.~\ref{slopes} and also refs.~\cite{HKM,DumRi,KPjp}.
Resonance decays are not included in the hydrodynamic spectra
on the $T=130$~MeV isotherm.

When going from SPS to RHIC energy, the model discussed here generally
yields only a slight increase of the inverse slopes,
although the specific entropy is larger by a factor of 4-5~! 
The reason for this behavior is the first-order phase transition that
leads to smaller transverse flow velocities than an ideal gas~\cite{soft}
(but to a larger expansion rate of the hadronization
volume~\cite{ExpSc}~!). For our initial
conditions, the collective transverse flow velocity
on the hadronization hypersurface increases only from $\approx0.3$
(Pb+Pb at SPS) to $\approx0.35$ (Au+Au at RHIC)~\cite{DumRi}.
As can be seen from the present calculation, this is not counterbalanced
by increased rescattering in the purely hadronic stage (compare to the
inverse slopes obtained from ``pure'' hydrodynamics with freeze-out on the
$T=130$~MeV isotherm~!).

In summary, we have employed relativistic hydrodynamics to describe the
evolution and hadronization of a hypothetical quark-gluon fluid
at CERN-SPS and BNL-RHIC energies. The produced hadrons are propagated within
a microscopic transport model (UrQMD). Interactions within the hadron gas
increase the collective flow beyond that present at hadronization, and
reduce the temperature below the QCD phase transition temperature (we assume
$T_C=160$~MeV).
As an exception, we find that multiple strange baryons practically do not
rescatter within the hadron gas. Their $m_T$-spectra are thus determined
by the conditions on the hadronization hypersurface, i.e.\ $T_C$ and
the collective flow created prior to hadronization.
Their spectra therefore are less sensitive to the
confined phase, $T<T_C$, but are closely related to the EoS of the
QGP and the phase transition temperature $T_C$. At RHIC energy, for all hadrons
$T^*$ or $\langle p_T\rangle$ are considerably smaller than predicted
by pure hydrodynamics with freeze-out on the $T=130$~MeV isotherm. Thus,
kinetic freeze-out occurs closer to the phase boundary than at SPS energy;
this is due to the different chemical composition of the central region
(more mesons, less baryons) and due to the larger expansion rate
of the hadronization volume.
\acknowledgements
We thank T.\ Ullrich and N.\ Xu for helpfull
comments.
A.D.\ gratefully acknowledges a fellowship by the German
Academic Exchange Service (DAAD).
S.A.B.\ is supported in part by the Alexander von Humboldt Foundation,
and by DOE grant DE-FG02-96ER40945.
M.B.\ thanks the Josef Buchmann Foundation for support. This work has been
supported in part by BMBF, DFG, and GSI.


\vspace*{-.8cm}
\begin{references}
\bibitem{QGP}
E.V. Shuryak, \Journal{\PR}{61}{71}{1980};
L. McLerran, \Journal{Rev. Mod. Phys.}{58}{1021}{1986};
B. M\"uller, \Journal{Rept. Prog. Phys.}{58}{611}{1995};
S.A. Bass, M. Gyulassy, H. St\"ocker, and W. Greiner,
\Journal{\JPG}{25}{R1}{1999}
\bibitem{StoePR} H. St\"ocker and W. Greiner, \Journal{\PR}{137}{277}{1986};
R.B. Clare and D. Strottman, \Journal{\PR}{141}{177}{1986}
\bibitem{relax}
P. Danielewicz, \Journal{\PL}{146}{168}{1984};
P. Levai and B. M\"uller, \Journal{\PRL}{67}{1519}{1991};
M. Prakash, M. Prakash, R. Venugopalan, and G. Welke,
\Journal{\PR}{227}{321}{1993}
\bibitem{microFO} L.V. Bravina, I.N. Mishustin, N.S. Amelin,
J.P. Bondorf, and L.P. Csernai, \Journal{\PL}{354}{196}{1995};
H. Sorge, \Journal{\PL}{373}{16}{1996};
S. Pratt and J. Murray, \Journal{\PRC}{57}{1907}{1998}
\bibitem{HKM} C.M. Hung and E. Shuryak, \Journal{\PRC}{57}{1891}{1998}
\bibitem{thOm} H. van Hecke, H. Sorge, and N. Xu,
\Journal{\PRL}{58}{5764}{1998}
\bibitem{expOm} 
E. Andersen et al.\ (WA97 Collaboration),
\Journal{\PL}{433}{209}{1998}
\bibitem{NA49chi}
H. Appelsh\"auser et al., (NA49 collaboration), 
\Journal{\PL}{444}{523}{1998}
\bibitem{Tmsyst} 
I.G. Bearden et al., (NA44 Collab.), \Journal{\PRL}{78}{2080}{1997}
\bibitem{UrQMD} S.A. Bass et al., \Journal{Prog. Part. Nucl. Phys.}
{41}{225}{1998}
\bibitem{ExpSc}
A. Dumitru, hep-ph/9905217
\bibitem{Bj}
J.D. Bjorken, \Journal{\PRD}{27}{140}{1983};
K. Kajantie and L. McLerran, \Journal{\NP}{B214}{261}{1983}
\bibitem{DumRi} A. Dumitru and D.H. Rischke, \Journal{\PRC}{59}{354}{1999}
\bibitem{NA49netB}
H. Appelsh\"auser et al., (NA49 collaboration),
\Journal{\PRL}{82}{2471}{1999}
\bibitem{BMS}
J. Letessier, A. Tounsi, U. Heinz, J. Sollfrank, and
J. Rafelski, \Journal{\PRL}{70}{3530}{1993};
P. Braun-Munzinger, J. Stachel, J.P. Wessels, and N. Xu,
\Journal{\PL}{365}{1}{1996};
C. Spieles, H. St\"ocker, and C. Greiner,
\Journal{\EPJ}{C2}{351}{1998};
M. Reiter, A. Dumitru, J. Brachmann, J.A. Maruhn,
H. St\"ocker, and W. Greiner, \Journal{\NP}{A643}{99}{1998}
\bibitem{T0} 
B. M\"uller and X.N. Wang, \Journal{\PRL}{68}{2437}{1992};
E. Shuryak, \Journal{\PRL}{68}{3270}{1992};
K. Geiger, \Journal{\PRD}{46}{4965}{1992};
T.S. Biro, E. van Doorn, B. M\"uller, M.H. Thoma, and
X.N. Wang, \Journal{\PRC}{48}{1275}{1993};
B. K\"ampfer and O.P. Pavlenko, \Journal{\ZPC}{62}{491}{1994};
K.J. Eskola and X.N. Wang, \Journal{\PRD}{49}{1284}{1994};
K.J. Eskola and K. Kajantie, \Journal{\ZPC}{75}{515}{1997}
\bibitem{microRHIC}
K. Geiger and J.I. Kapusta, \Journal{\PRD}{47}{4905}{1993};
T. Sch\"onfeld, H. St\"ocker, W. Greiner, and H. Sorge,
\Journal{\MPLA}{8}{2631}{1993};
L. Gerland et al., nucl-th/9512032;
S.E. Vance, M. Gyulassy, and X.N. Wang, \Journal{\NP}{A638}{395c}{1998}
\bibitem{CF}
F. Cooper and G. Frye, \Journal{\PRD}{10}{186}{1974}
\bibitem{NA49plmi}
P.G. Jones and the NA49 collaboration, \Journal{\NP}{A610}{188c}{1996};
\bibitem{NA49L}
C. Bormann et al., (NA49 collaboration), \Journal{\JPG}{23}{1817}{1997}
\bibitem{PDg} C. Caso et al., \Journal{\EPJ}{C3}{1}{1998}
\bibitem{BassDu} S.A. Bass et al., nucl-th/9902062
\bibitem{KPjp}
H. St\"ocker, A.A. Ogloblin, and W. Greiner,
\Journal{\ZPA}{303}{259}{1981};
T. Cs\"org\"o and B. L\"orstad, \Journal{\PRC}{54}{1390}{1996};
B. K\"ampfer, O.P. Pavlenko, A. Peshier, M. Hentschel, and G. Soff,
\Journal{\JPG}{23}{2001}{1997}
\bibitem{soft}
M. Kataja, P.V. Ruuskanen, L.D. McLerran, and H. von Gersdorff,
\Journal{\PRD}{34}{2755}{1986};
C.M. Hung and E. Shuryak, \Journal{\PRL}{75}{4003}{1995};
D.H. Rischke and M. Gyulassy, \Journal{\NP}{A597}{701}{1996};
\Journal{\NP}{A608}{479}{1996}
\end{references}
\end{document}